# AI-Powered Semantic Segmentation and Fluid Volume Calculation of Lung CT images in Covid-19 Patients

*Sabeerali K.P[1], Saleena T.S[2], Dr.Muhamed Ilyas P.[3], Dr. Neha Mohan, MD[4]

[1] Sullamussalam Science College Areekode,India

[1]*sabeeralikp233@gmail.com

ORCID : 0000-0001-5184-2054

[2] Sullamussalam Science College Areekode,India

[2]tssaleena@gmail.com

ORCID : 0000-0001-5709-355X

Sullamussalam Science College Areekode,India

muhamed.ilyas@gmail.com

Govt. Medical College, Manjeri, India

drnehamohan@gmail.com

**Abstract.** COVID-19 pandemic is a deadly disease spreading very fast. People with the confronted immune system are susceptible to many health conditions. A highly significant condition is pneumonia, which is found to be the cause of death in the majority of patients. The main purpose of this study is to find the volume of GGO and consolidation of a covid-19 patient so that the physicians can prioritize the patients. Here we used transfer learning techniques for segmentation of lung CTs with the latest libraries and techniques which reduces training time and increases the accuracy of the AI Model. This system is trained with DeepLabV3+ network architecture and model Resnet50 with Imagenet weights. We used different augmentation techniques like Gaussian Noise, Horizontal shift, color variation, etc to get to the result. Intersection over Union($IoU$) is used as the performance metrics. The IoU of lung masks is predicted as 99.78% and that of infected masks is as 89.01%. Our work effectively measures the volume of infected region by calculating the volume of infected and lung mask region of the patients.

.

## 1  Introduction

The corona virus disease 2019 (COVID-19) is an ongoing pandemic, gripping the world over and affecting millions of people ever since the first outbreak in December 2019 at Wuhan, China. It is an infectious disease caused by the novel RNA virus SARS-CoV-2 (severe acute respiratory syndrome coronavirus-2) mainly manifesting as respiratory illness of varying severity. Though the disease can be asymptomatic in healthy individuals, it can cause significant morbidity in others, especially in elderly patients and those with other co-morbidities where it can even be fatal.

As per WHO statistics, globally, there have been 11,36,95,296 confirmed cases of COVID-19 till March 1 2021, including 25,26,007 deaths [1]. The rapid spread of COVID-19 to global level pandemic created the urgent need for a reliable and efficient way of diagnosing patients. Even though a positive RT-PCR test is required for definitive diagnosis of COVID-19, CT Chest has a potential role in diagnosis, detection of complications, and most importantly prognostication of the disease. The presence and extent of abnormality in the lungs on CT is based on the stage and severity of the disease and the most common abnormal findings are bilateral peripheral ground-glass opacity and/ or consolidation with predilection for lower lobes of lungs [2]. Ground glass opacity and consolidation are areas of increased attenuation/ density in lungs on CT and represent infected/ inflamed lungs in patients with COVID-19 pneumonia. Detecting these findings and assessing the severity of the involvement of lungs will help triage patients properly so that the worst affected ones are quickly identified and addressed, leading to better patient care and treatment.

Usually, radiologists qualitatively evaluate the extent of lung volume infected in CT and issue reports. Manual CT image segmentation takes time and may have different hurdles like variation in shapes of ROI, difficulty in edge detection of ROI, clarity of image, noise inbuilt in the medical imaging devices etc[4]. Also, the ratio of the number of radiologists available to report to the number of CT images to be read is very much on the lower side. And as the number of patients increases exponentially, it further becomes a herculean task, affecting workflow. A fast auto-contouring computerized tool to accurately quantify the infection lung regions in COVID-19 infection will be a boon in this testing time and is a need of the hour. So we come up with a method to help radiologists in which a Computer System can segment the lung CTs and quantify infected lung volumes so that health care providers can prioritize patients who need critical care and can make the right treatment decisions. This will be done based on the severity of the disease which can be categorized using this system.

[15] The semantic segmentation using Fully Convolutional Neural Networks may end up with fuzzy object boundaries and low-resolution images as it causes loss of information due to convolution and pooling. This has been overtaken by the DeepLab series that uses Atrous Spatial Pyramid Pooling. Among them, DeepLabV3 version onwards, the required features are extracted from the pre-trained networks like VGG, ResNet, DenseNet etc. The proposed system uses ResNet50 as the backbone network along with DeepLabV3+ which is an extended version of DeepLabV3[11]. The percentage of infected regions can be calculated from the total lung area.

Percentage of the infected portion = (volume of infection mask/volume of total lung mask) *100

This system can be used in collaboration with the CT scanner and set a threshold value so that patients can be prioritized based on that value. The radiologist will be notified of this value and such patients will be treated with more care and attention and others will be discharged. This helps to utilize the hospital resources efficiently.

## 2    Related Works

Tamer F. Ali et al.[16] narrates the findings that a radiologist should identify from lung CT images in the case of Covid-19. The key infection indicators are ground-glass opacity (GGO) and consolidation.Feng Shi et al.[3], made a sail on different AI techniques that applied in X-Ray and CT of Covid19 patients and they made a consolidation about AI-empowered contactless medical image acquisition workflows, segmentation, diagnosis, and follow-up studies. In all these cases, U-Net architecture is predominant. The laboratory tests, especially the RT-PCR test is now considered as the gold standard for the diagnosis of covid19. But still, it may be inadequate in some situations. Ophir Gozes et al.[9] developed an AI-based automated CT image analysis tool that classifies the Covid affected patients using thoracic CT and tracks the disease burden. The work showed 98.2% sensitivity, 92.2% specificity on datasets of Chinese control and infected patients. Shuai Wang et al.[10]  used inception migration-learning model to set a deep learning model. The performance metrics of accuracy, specificity, and sensitivity of internal validation are 82.9%, 80.5%, and 84% respectively whereas the same of external testing is 73.1%, 67%, and 74%. [13]The DeepLab has been introduced by Liang-Chieh Chen et al. which was the state-of-art network in the competition of semantic image segmentation task using PASCAL VOC-2012. Its mIOU value has been measured as 79.7%. [11]A novel model DeepLabV3+ which is the extension of DeepLabV3 has been introduced by Liang-Chieh Chen in which DeepLabV3 acts as the encoder module and an additional decoder module to refine the segmentation process. This architecture along with ResNet-101 and Xception as network backbone are used in which the Xception model results the best test set performance of 89%.

## 3        Methods

### 3.1        Data    preparation

Images have been downloaded from the site *radiopedia.org* and *coronacases.org*. This work has been implemented using the pytorch library. The Med2image library converts each slice of Nifti image and its mask into a corresponding .png file with the same dimension without any loss of data. The system used 512 X 512 X 3 and 512 X 512 images and masks respectively. Pre-processing is not explicitly giving for this network, as it makes use of the same of the pre-trained model. Each pre-trained model will have its pre-processing steps. We just have to input the model name and its corresponding weight obtained from ImageNet, which will give pre-processed data as output.

The whole dataset has been divided into 3 sets: training set-1789 images and corresponding masks(82%), validation set – 207(10%) testing set – 176 images(8%)

### 3.2 Data Augmentation

[8]As the deep neural networks heavily rely on big data for better performance and medical image analysis has no access to the same data augmentation is an inevitable factor in such cases. The augmentation using keras will change the pixel value of the image. But the semantic segmentation highly relies on the pixel and it should not be changed during augmentation. In such cases, **Albumentation** is the right choice. In this system, the training data has been augmented using the this library**.** Even though the library provides more than 70 augmentations, this system used horizontal flip, shift scale rotate, random crop, additive Gaussian noise, perspective, CLAHE, random brightness, sharpen, blur, motion blur, random contrast, and hue saturation.

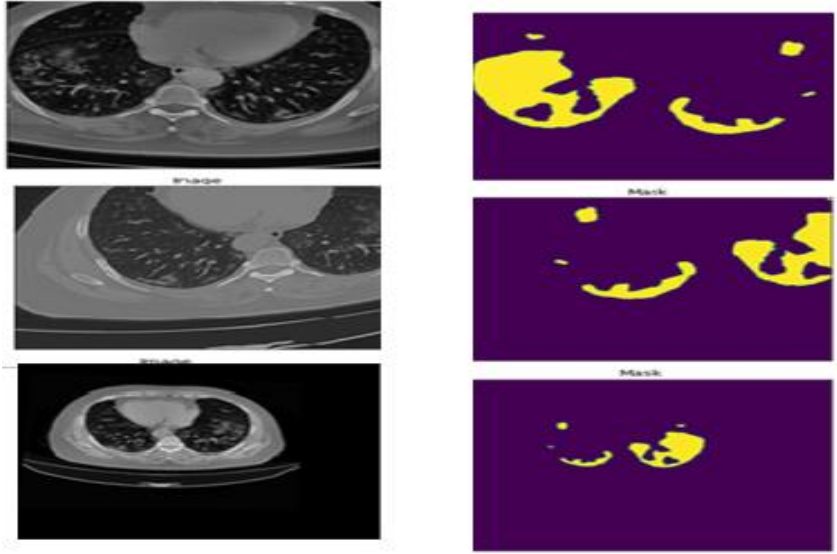

**Fig. 1.** Different augmentations done on a single CT image (left) and its masks at right

### 3.3 Semantic Segmentation

Segmentation is a process in which it not only identifies whether the disease present or not but also contours the area which is affected by the disease. It helps in the localization of disease and quantization of the volume of disease. Whereas semantic segmentation aims at the pixel-wise labeling of the image using the corresponding category to which it belongs [11][17][18]. Each pixel of the image is checking whether it belongs to fluid content or not. The segmentation task can be performed without explicitly coding with the help of segmentation models that provide pre-configured models and backbones. This backbone refers to any pre-trained classification model without dense layer, which is used for feature extraction to build the model. Figure(2) depicts the workflow of the system, that uses DeepLabV3+ architecture with ResNet50 as the feature provider.

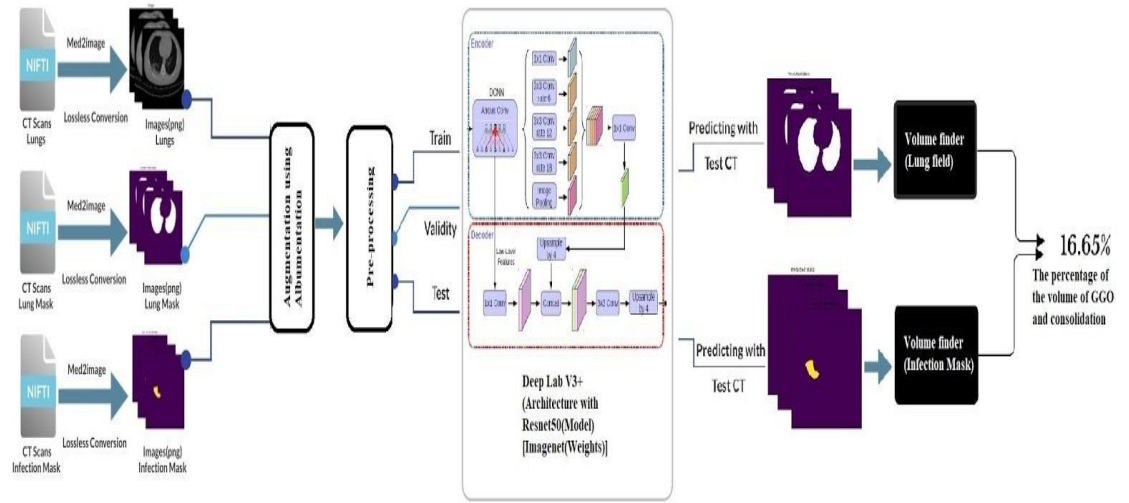

**Fig. 2.** Segmentation model using DeepLabV3+ and ResNet50 as backbone is finding the volume of affected part in lung CT

**DeepLabV3+ architecture:** [12]DeepLabV3+is the latest and the most effective architecture in the DeepLab series which is the invention of Google. The Atrous Spatial Pyramid Pooling or ASPP and the encoder-decoder architecture make this version capable of outperforming all other similar kinds. This makes it possible to create the output image of same size as the input image, as here occurs the pixel to pixel mapping in semantic segmentation.[11]This combination is leading to semantic image segmentation tasks. [13]ASPP enables the object and image context segmentation in multiple scales. [14]The encoder-decoder architecture consists of a contracting path that extracts the required features and an expanding path that will localize the affected region. Liang-Chieh Chen et al. [11] depicts the above-described features of DeepLabV3+ through figure(3).

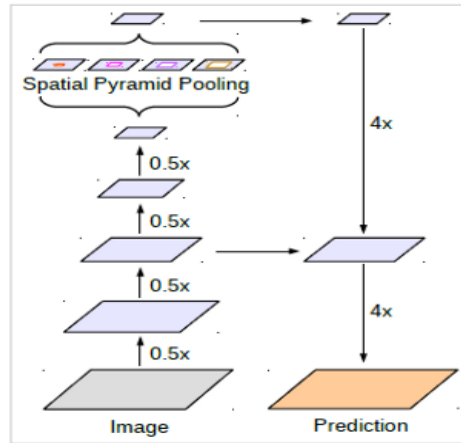

**Fig. 3.** Architecture of DeepLabV3+

**ResNet-50 as feature extractor pre-trained network:** ResNet-50 is used here as a network backbone for feature extraction. ResNet is one of the powerful classification networks that proved its excellence in the ***ILSVRC 2015*** classification challenge. It is pre-trained on the ImageNet dataset. So we can load the pre-defined weights of the model, freeze the encoder part as it is, and need to begin with the decoder part only while training. [20]It has 48 convolution layer, 1 max-pooling, and 1 average pooling layer. [21] The figure shows the residual block of the deep residual network.

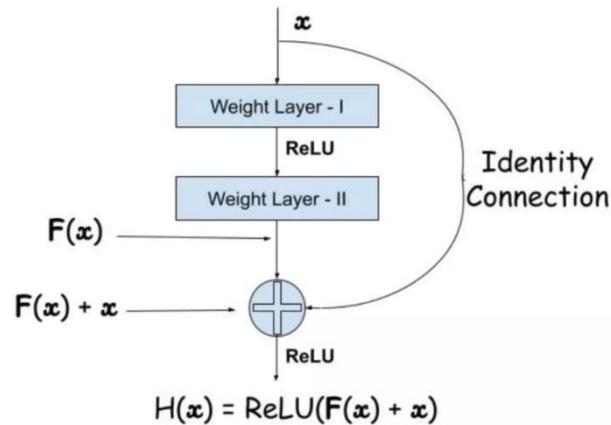

**Fig. 4.** Residual Function block of ResNet

## 4      Discussion

This study demonstrates that how accurately the GGO and consolidation in CT scan images has been segmented using DeepLabV3+ with the model of ResNet50. The Nifti images of original CT scan, lung mask and infection mask has been losselessly converted to .png images and after augmentation and pre-processing they are fed into the DeepLabV3+ architecture with model ResNet50 which using Imagenet weights. The trained model is saved and testing and validation is performed based on that model.

The accuracy is not a good metric in the case of semantic segmentation. Because, in some cases the entire background may be matching but the mask area may not be matching very well. So among the several performance metrics to evaluate the efficiency of the model, the Intersection over Union or IoU is used in this scenario. [21]It compares the predicted segmented mask with the corresponding ground truth. The IoU of lung mask is predicted as 99.78% and IoU of infected mask is as 89.01%. The percentage of infected region is calculated from the total lung area. The volume of total lung and that of segmented mask is measured separately further part done using the equation,

Percentage of the infected portion = (volume of infection mask/volume of total lung mask) *100

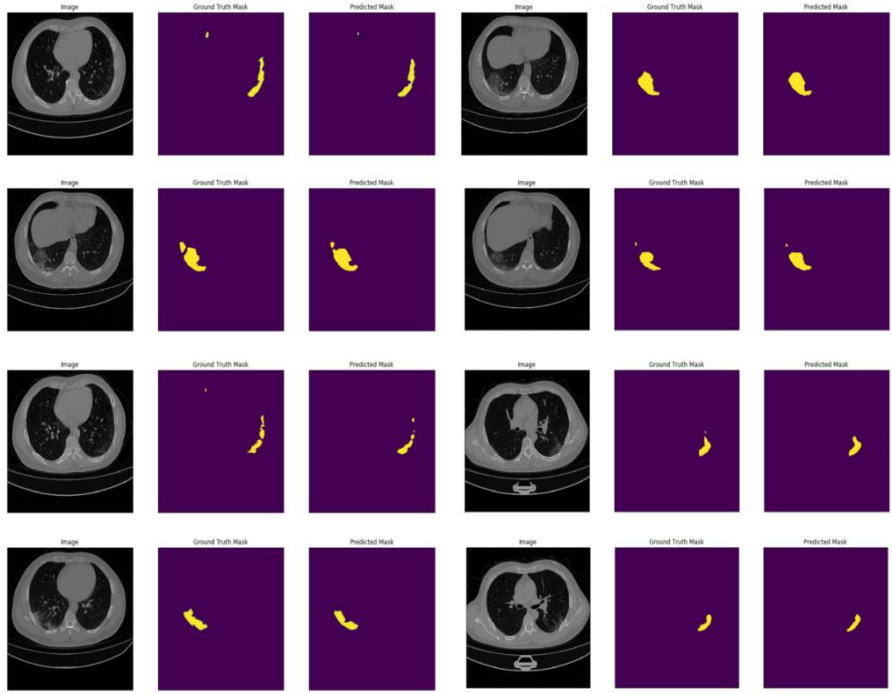

**Fig. 5.** CT image(from left), Ground truth mask and Predicted mask(output of the system)

## 5      Conclusion

This study proves that deep learning is an effective tool to help radiologists in segmenting and volume finding of GGO and consolidation in COVID-19 patients. This can reduce the inter-observer variability and subjectivity of the radiologist. This will also greatly reduce their workload and reporting time. This system can prioritize the patients based on the volume of infection which can ensure that medical care can be given to those who are really in need.